\documentclass[a4paper,fleqn,usenatbib]{mnras}

\usepackage[T1]{fontenc}
\usepackage{ae,aecompl}

\usepackage{graphicx}	
\usepackage{amsmath}	
\usepackage{amssymb}	

\title[Dynamical history of a binary cluster: Abell 3653]{Dynamical history of a binary cluster: Abell 3653}

\author[T. Caglar \& M. Hudaverdi]{
Turgay Caglar$^{1}$ and Murat Hudaverdi$^{1, 2}$\thanks{E-mail: hudaverd@yildiz.edu.tr (MH)}
\\
$^{1}$Y{\i}ld{\i}z Technical University, Faculty of Science and Art, Department of Physics, Istanbul 34220, Turkey\\
$^{2}$AUM, College of Engineering and Technology, Department of Science, Dasman 15453, Kuwait}

\date{Accepted  2017 August 10. Received 2017 August 10; in original form 2017 January 13}

\pubyear{2017}
   \volume{472}

\begin{document}
\label{firstpage}
\pagerange{2633--2642}
\maketitle

\begin{abstract}
We study the dynamical structure of a bimodal galaxy cluster Abell 3653 at z=0.1089 by using combined optical and X-ray data. 
Observations include archival data from Anglo-Australian Telescope and X-ray observatories of XMM-Newton and Chandra. 
We draw a global picture for A3653 using galaxy density, X-ray luminosity and temperature maps. 
Galaxy distribution has a regular morphological shape at the 3 Mpc size. 
Galaxy density map shows an elongation EW direction, which perfectly aligns with the extended diffuse X-ray emission.
We detect two dominant grouping around two brightest cluster galaxies (BCGs).
The BCG1 (z=0.1099) can be associated with the main cluster A3653E, 
and a foreground subcluster A3563W concentrated at the BCG2 (z=0.1075).
Both X-ray peaks are dislocated from BCGs by ($\sim$35 kpc), which suggest an ongoing merger process. 
We measure the subclusters' gas temperatures 4.67 and 3.66 keV, respectively. 
Two-body dynamical analysis shows that A3653E \& A3653W are very likely (93.5\% probability) gravitationally bound. 
The highly favoured scenario suggests that two subclusters with the mass ratio of 1.4 are colliding close to the plane of sky ($\alpha$=17$^\circ$.61) with 2400 km $s^{-1}$, 
and will undergo core passage in 380 Myr.
Temperature map also significantly shows a shock-heated gas (6.16 keV) in between the subclusters, which confirms the supersonic infalling  scenario.  

\end{abstract}





\section{Introduction}
Cosmological models predict that small structures form first then progressively merge into larger structures. 
Clusters of galaxies are the largest gravitational entities of the Universe and their formation occurs at early epochs (z $\gtrsim$ 1).   
Observations of forming clusters at high redshifts require long exposures and involve crucial technical difficulties.
Nearby systems can still provide critical information about the structural formation; however, the evolutionary effects should be considered cautiously.

The dynamical consequence of a cluster merger is the transfer of energy ($\gtrsim$10$^{64}$ ergs s$^{-1}$) and angular momentum of the merging subclusters to the system. 
Collisional nature of hot plasma and non-collisional nature of galaxies result in different reaction time scales during a merger \citep[e.g.,][]{WF1995, Roettiger1997}.
In such merging conditions extended ICM is strongly effected while individual galaxies can pass through. 
Brightest Cluster Galaxy (BCG) is defined as brightest galaxy in the cluster. 
BCGs are generally elliptical and are expected to be sitting at the bottom of the potential well of the parent cluster \cite[e.g.,][]{QL1982,adami98,adami2000}. 
The merger of BCG hosting clusters creates velocity offsets and dislocates BCG from the dynamical centre. 
Several observations have identified displaced BCGs from the cluster centre \cite[e.g.,][]{SM2005,SH2010}
and the number of BCGs with significant velocity gradient from the cluster mean \cite[see e.g.,][]{Beers1991}.
Velocity comparison can measure the system's deviation from a relaxed configuration. 
Density maps which are produced by position and velocity information of cluster member galaxies,
also reveal the footprints of substructures \cite[e.g.,][]{quin1996,berr2009,shak2016}.

With the recent improvements on high angular resolution cameras, X-ray observations have played a key role to investigate merging clusters.
The fate of the merger energy and the complexity of collisional ICM have been considerably settled. 
Apparently, kinetic energy of the merging structures is converted into thermal energy of the plasma, 
which has been observed in many clusters \cite[e.g.,][]{mv2007,bm2008}.

\begin{figure*}
 \begin{center}
 \includegraphics[width=8.5cm]{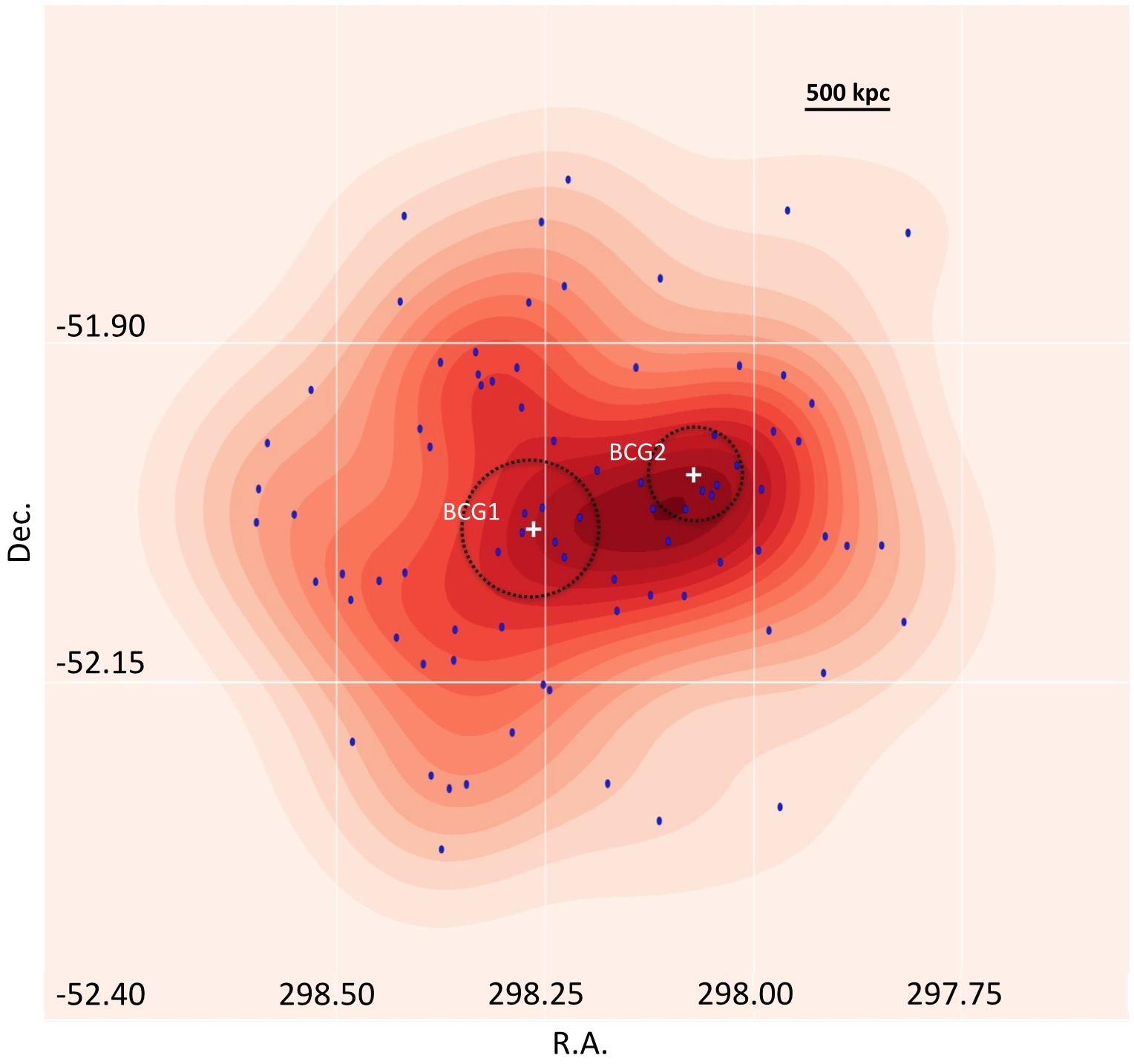}
  \includegraphics[width=7.6cm]{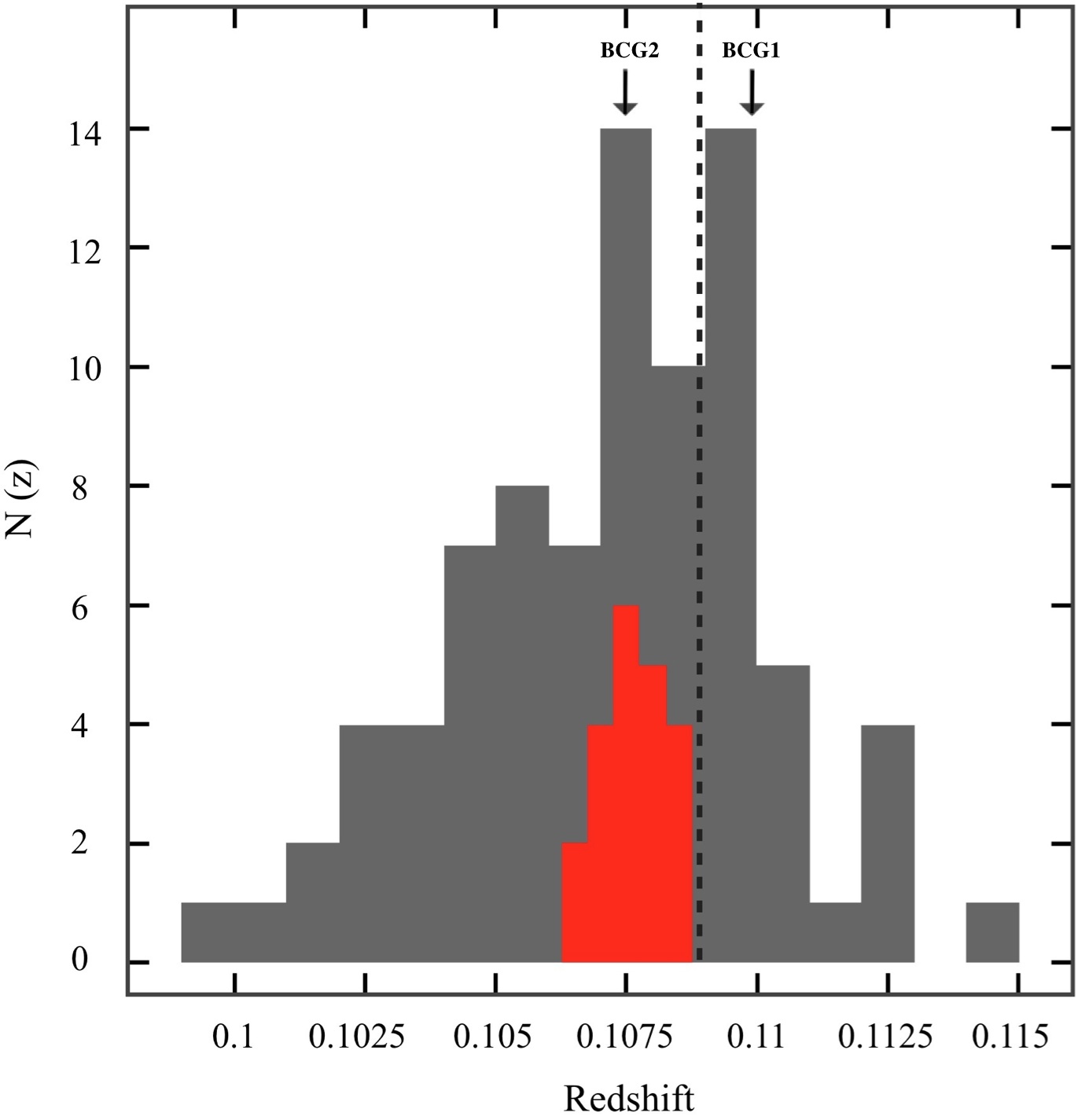}
\caption{\label{opt_density} Left: Galaxy density map A3653  within 1.5 Mpc radius field. The white crosses indicate the BCG positions.
Right: Redshift histogram of A3653 for 84 member galaxies. 
The vertical dotted line gives the mean cluster value, and two arrows are the redshifts of the BCGs.
The red histogram shows the member galaxies around the BCG2 ($0.10625<z<0.10875$).}
\end{center} \end{figure*}

In this paper, we present the structural analysis results of Abell 3653 (A3653 hereafter). 
The cluster's BCG is known to have one of the most extreme radial velocities \citep{Pimbbett}, 
which is a strong suggestion for a recent or ongoing merger activity. 
The original optical data were studied for signs of substructure by \citet{Pimbbett}, but they did not report any significant grouping.
A3653 was also studied in the \textit{MCMX} meta-catalogue based on \textit{ROSAT} All Sky Survey \citep{Piffaretti}.
The \textit{MCMX} provides a redshift of 0.1069,  a standardized $0.1-2.4$ keV band luminosity $L_{500}=1.57\times10^{44}$ ergs s$^{-1}$, 
total mass of $M_{500}=2.5\times10^{14}$ $M_{\odot}$, and radius of $R_{500}=925.8$ kpc. 
In order to understand the dynamical structure of A3653, 
we investigate X-ray morphology and temperature distribution with archival $XMM$-$Newton$ and $Chandra$ observations.  
Raw X-ray image shows a clear bimodal structure with X-ray centroids:  
A3653E (J2000, RA:19$^{h}$53$^{m}$01.9$^{s}$, DEC.:$-52^{d}$02$^{m}$13$^{s}$), and A3653W (J2000, RA: 19$^{h}$52$^{m}$17.3$^{s}$, DEC.:$-51^{d}$59$^{m}$50$^{s}$) (see Figure-\ref{raw_image}).

In this study, we aim to describe X-ray morphology and temperature structure of Abell 3653. 
Besides, we focus on explaining A3653 physical mechanism occurred with dynamical merge activities. 
We compare our X-ray results with optical results studied by \citet{Pimbbett}. 
The optical study signs that A3653 has 111 galaxies at cluster field. 
A3653 is located at RA= 19$^{h}$53$^{m}$00.9$^{s}$ and Dec.: -52$^{d}$01$^{m}$51$^{s}$ with redshift (z) value of 0.1089 \citep{struble}. 
In addition, A3653 is a rich centrally condensed cluster \citep[]{Greg}. 

This paper is organised as follows; in Section 2,  we describe the data used in our analysis. 
In Section 3, the analysis of the X-ray and optical observations is described, the temperature distribution is studied in detail. 
In Section 4, we discuss cluster dynamics and possible subclusters. 
We summarise our results in Section 5.
We adopt cosmological parameters $H_{0}=70\ \mathrm{km}\,\ \mathrm{s}\,^{-1}\ \mathrm{Mpc}\,^{-1}$, $\Omega_{M}$=0.27 and $\Omega_{\Lambda }$=0.73 in a flat universe. 
For this cosmology, an angular size of 1 arcmin corresponds to a physical scale of 119.22 kpc  at the redshift of A3653 (z=0.1089).
Unless otherwise stated, reported errors correspond to 90\% confidence intervals.


\section{Observations and Data Reduction}

\subsection{Optical Spectroscopic Data}

We obtained galaxy spectroscopic redshifts using Anglo-Australian Telescope (\textit{AAT}). 
In the literature, there are several velocity measurement studies performed for a variety of samples. 
We selected galaxies within a 3.0 Mpc radius of the A3653 centroid with a spectroscopic redshift falling in the interval $0.099 < z_{spec} < 0.115$,
which fairly captures the range of galaxies associated with A3653.
Our selection includes total 87 members; 83 galaxies from 2dF \citep{Pimbbett} and 4 galaxies from 6dF Galaxy Survey final redshift release \citep{jones2009}.
The member galaxies are listed in the appendix.

Projected galaxy density map of member galaxies is shown in Figure-\ref{opt_density} left-panel. 
Galaxies identified as cluster members from spectroscopy are marked with dots. 
Two plus signs are the locations of the BCGs. 
Two circles are superimposed for visual aid, shows the extended X-ray count extraction regions used for the spectral analysis in the following sections.
The galaxy density distribution shows an elongated structure East-West axis with a little tilt. 

\begin{figure*}
 \begin{center}
 \includegraphics[width=16cm]{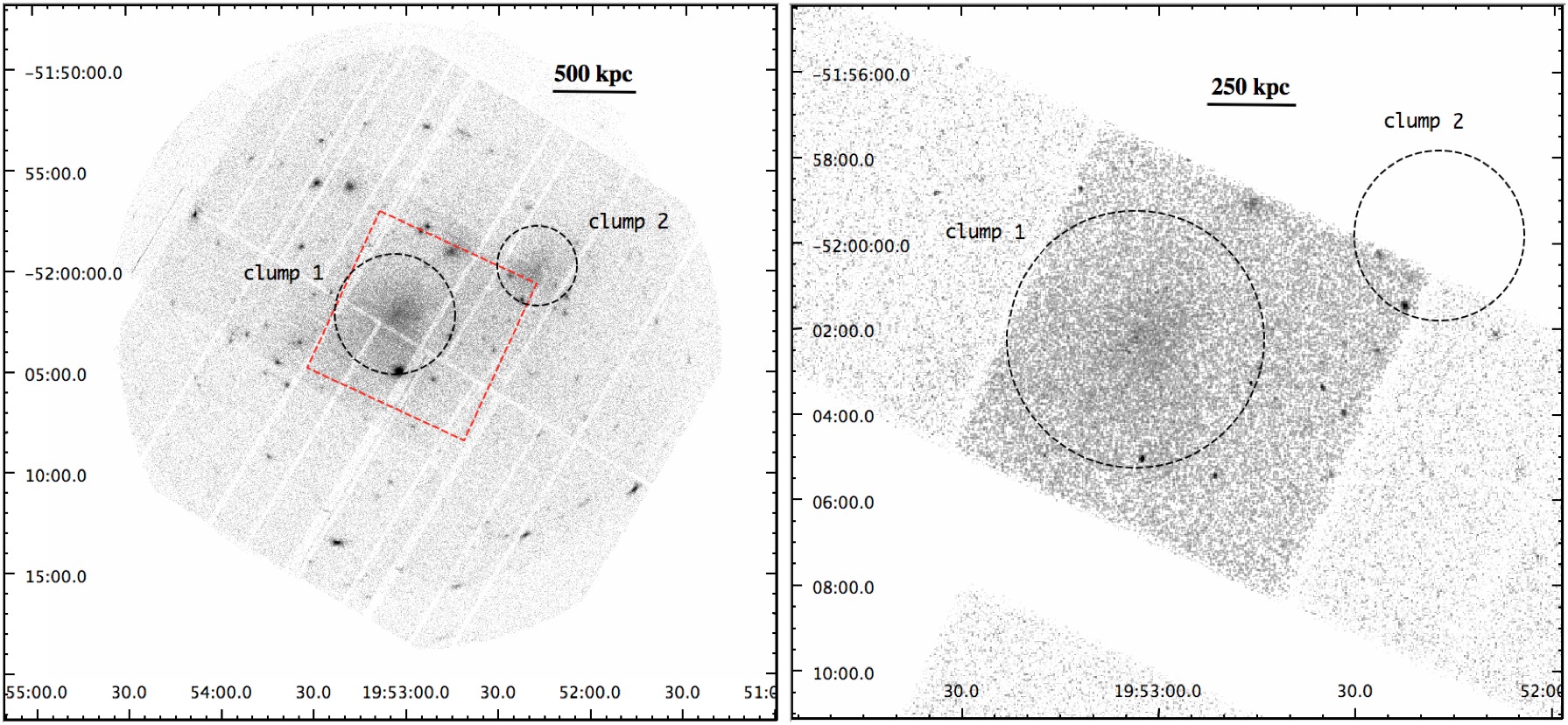}
 \caption{\label{raw_image}
The panels are the \textit{XMM-Newton} mosaic image of MOSs/PN in the 0.3-10 keV (left) and the $Chandra$ ACIS-S3 in the 0.5-7 keV (right) energy ranges,
background and exposure corrected images. 
The subclusters are circled with the circles used for the spectral extraction.
The tilted square region on the \textit{XMM-Newton} image corresponds \textit{Chandra} FOV.
The images are adaptively smoothed in order to highlight the structures with extended plasma emission.}
\end{center} 
\end{figure*}

On the basis of 87 redshifts, we plot a redshift histogram of A3653 (see Figure \ref{opt_density} right-panel). 
The vertical dotted line shows the mean cluster redshift value of $z=0.1089$.
The cluster's BCG is known as having one of the most extreme radial velocity $\sigma_{cz}=683\pm96$ km $s^{-1}$ 
away from the mean cluster velocity \citep{Pimbbett}, which locates it considerably further at $z=0.1099$.
Although a gaussianity is visible around the cluster mean $z=0.1089$, 
deviations from a single Gaussian can be attributed to the subclusters around two BCGs.  
The BCG1 at $z=0.1099$ can be associated with the main cluster, 
and a foreground west clump concentrated around the galaxy 2MASX J19521735-5159465 at z=0.1075 \citep{jones2009}.
The galaxy is catalogued with ($r_F$,$b_J$) = (15.20,16.11) magnitudes in the NASA Extragalactic Database (NED). 
Considering its location at the X-ray centroid of A3653W 
and being two magnitude brighter than neighboring members, 
2MASX J19521735-5159465 is named as BCG2 hereafter.
The BCG2 was unexpectedly not included in the 2dF galaxy survey study of \citet{Pimbbett}.
This is possibly the reason why the cD galaxy of A3653 and its peculiar velocity could not be explained with no significant grouping.
The redshift values of the two BCGs are indicated in the histogram Figure-\ref{opt_density} right-panel.  
Since we are interested in the potential subgroups associated with merging scenarios, 
we also identify the galaxies clustering around BCG2 
with $V_r$ = 32221$\pm$350 km s$^{-1}$ interval ($0.10625<z<0.10875$) displayed in Figure \ref{opt_density} (red colour).
Their spatial properties are discussed with X-ray analysis results in later sections, consecutively.



\begin{table}
\begin{center}
\caption{Log of X-ray Observations.}
\begin{tabular}{@{}cccl@{}} 
\hline
\hline
ObsID	    	& 	Satellite		& 	Date Obs		& Effective Exposure   	\\
            		&                   		&               		& Time (ks)          	 	\\
\hline
10460       	& \textit{Chandra}  	& 2009-07-05    	& ACIS 43.6 (\%98.6)       	\\
0691820101  	&\textit{XMM-Newton}& 2013-03-18    	& MOS1 48.6 (\%75.2)	\\
			& 				& 				& MOS2 48.8 (\%75.5)	\\
			& 				& 				& PN 47.5 (\%78.3)	     	\\
\hline
\end{tabular}
\label{obslog}
\end{center}
\end{table}

\subsection{X-Ray Observations}

We used \textit{XMM-Newton} and \textit{Chandra} archival data for our analysis (see Table-\ref{obslog}).
\textit{XMM-Newton} observation was performed March 19, 2013 for an exposure time of 64 ks with observation id 0691820101.
The MOS and PN detectors in Full Frame Mode and Extended Full Frame respectively, with the thin filter. 
 Observation data was obtained from XMM-Newton Science Archive. 
We performed the data processing and background modeling with the XMM Extended Source Analysis Software ($\it{XMM}$-$\it{ESAS}$: \cite{SN1}) 
and \textit{XMM-Newton} Scientific Analysis System (\textit{XMMSAS}) v14.0.
The event files for MOSs and pn were created using \texttt{emchain} and \texttt{epchain}, respectively. 
We filtered bad pixels, bad columns and cosmic rays using \texttt{evselect}. 
The total filtered MOS1, MOS2 and pn exposure times are 48.6 ks, 48.8 ks and 47.5 ks, respectively. 

The \textit{Chandra} observation of A3653 was carried out on 2009 July 5 (Obs ID 10460) for a total exposure of 44.2 ks 
with CCDs 3, 5, 6, 7, and 8 of Advanced CCD Imaging Spectrometer (ACIS) in operation, telemetered in the VFAINT mode. 
We obtained observation data from Chandra Data Archive. 
Most of the X-ray emissions of A3653 are covered by the AICS-S3 chip. 
Therefore, this study focuses on the S3 chip.
The data were reprocessed from level 1 event files using CIAO 4.8 and CALDB 4.7.0.
The total altered ACIS exposure time is 43.6 ks.

\begin{figure*}
 \begin{center}
 \includegraphics[width=16cm]{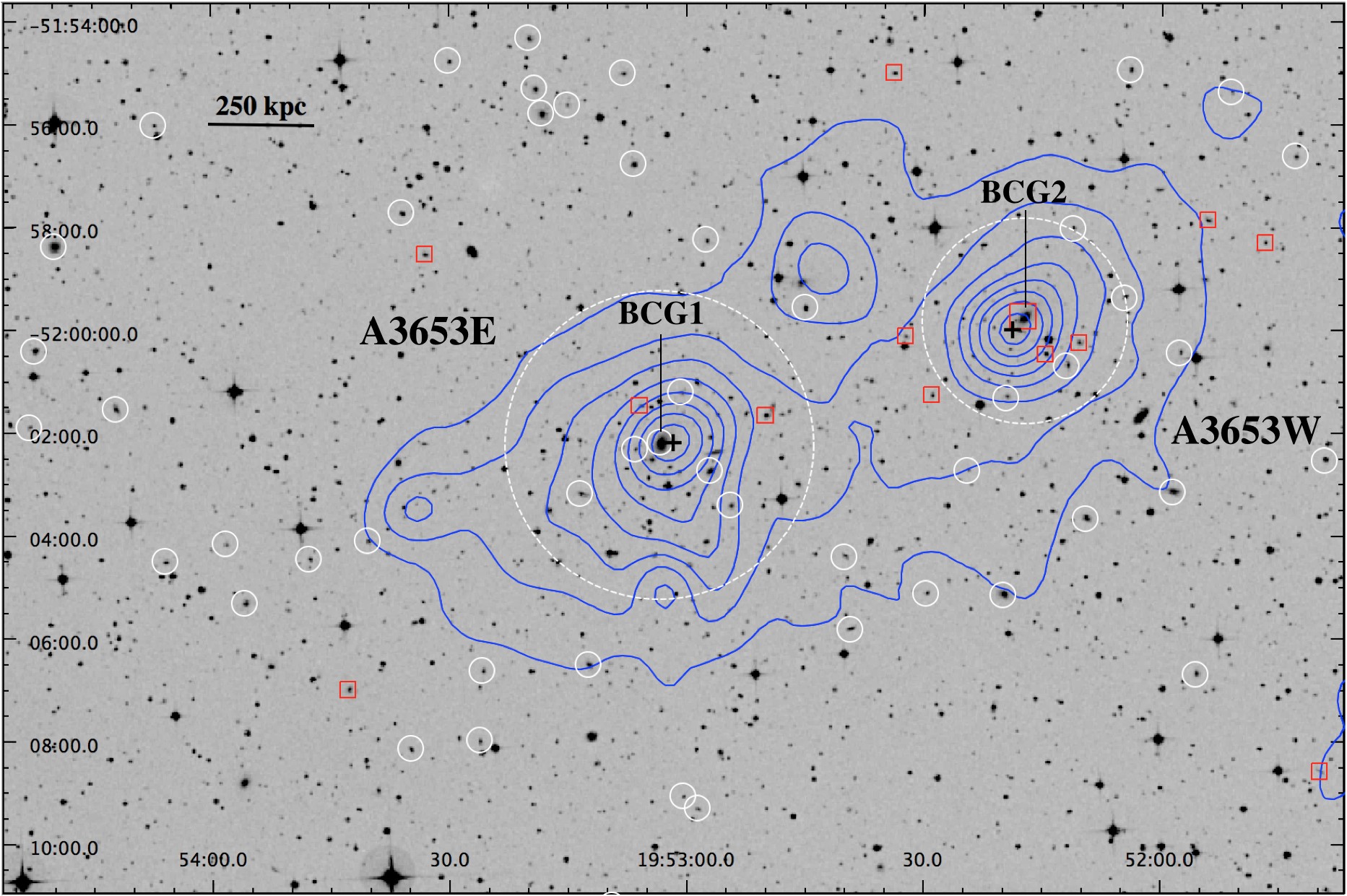}
 \caption{\label{optic} 
Contours of the X-ray diffuse emission at $0.3-10$ keV from MOS logarithmically spaced by a factor of 0.7 and overlaid on the DSS optical image.
The subclusters are indicated with dashed circles. 
The positions of the subcluster X-ray peaks are indicated by plus signs,
and associated BCGs are pointed by arrows.  
The galaxies identified as cluster members are marked with the white circles. 
The red boxes indicate the galaxies with $V_r$ = 32221$\pm$350 km s$^{-1}$ interval around BCG2.}
\end{center} \end{figure*}

\section{Analysis}

We generated broad-band images from \textit{XMM-Newton} in the $0.3-10$ keV and \textit{Chandra} in the $0.5-7$ keV, respectively. 
Figure \ref{raw_image} shows adaptively smoothed images without exposure and background correction. 
The left-panel is \textit{XMM-Newton} mosaic image MOSs/PN.  
Exposure corrected \textit{Chandra} image of extended diffuse emission for A3653 is created by using \texttt{fluximage} (Fig. \ref{raw_image} right-panel).
\textit{XMM-Newton} observation covers almost entire cluster emission, while narrow \textit{Chandra} field-of-view (FOV) of ACIS-S3 only collects photons from the main cluster. 
The tilted square at the \textit{XMM-Newton} image displays ACIS-S3 FOV, and the two circles are selected X-ray emission regions for global spectral analysis. 
The subclusters do not have bright central cores or regular morphologies, as is typical of non-cool core clusters.

The locations of the point sources in the FOV are detected using \texttt{edetect\_chain} in the $0.2-12$ keV energy band for \textit{XMM-Newton}. 
For \textit{Chandra} data, we used CIAO's \texttt{wavdetect} in the $0.5-8$ keV energy band. 
We have all point sources at 4$\sigma$ confidence level excluded by masking the event files and 
thereby decreasing the contribution of point sources to the extend ICM emission.

Figure \ref{optic} shows the contours from the combined MOS images in the $0.3-10$ keV energy band, overlaid on the DSS optical image for A3653. 
Since the PN image is disrupted by chip gaps at the central regions (see Fig. \ref{raw_image}), we rejected PN data for imaging and the related analysis. 
The X-ray image was adaptively smoothed and corrected for the background, exposure time, and vignetting.
Two regions used for the spectral extraction for the subcluster extend emission are indicated with dashed circles. 
The positions of the subcluster X-ray peaks (A3653E \& A3563W) are indicated by plus signs. 
BCG1 and BCG2 are pointed by arrows to underline their evident relations with A3653E and A3653W, respectively. 
However the X-ray peaks does not exactly match with the BCGs, there is a clear shift $\sim$35 kpc for both. 
These type of positional disturbances can be associated with an ongoing merger activity, which may dislocate BCGs from the dynamical centre. 
The cluster member galaxies from spectroscopy are marked with circles. 
The red boxes are the members of the sub-sample, which has redshift values clustering at A3653W redshift ($z=0.1075\pm0.0012$).
At this redshift range there are 21 members but spatially dispersed, which is not straightforward to determine an optical subcluster. 


\subsection{Spectral Analysis}

We examined the spectra of each subcluster to determine the global properties of the ICM.
The spectrum of A3653E is extracted and studied for \textit{XMM-Newton} and \textit{Chandra} since it is clearly viewed from both cameras. 
The western sub-cluster A3563W is analysed only by \textit{XMM-Newton} for lack of \textit{Chandra} observation in that region. 
\textit{XMM-Newton} spectra and response files were generated using \texttt{evselect} V3.62, \texttt{rmfgen} V2.2.1 and \texttt{arfgen} V1.90.4.
An annular region $11^{\prime}-12^{\prime}$ away from the cluster centroid is used to extract the local background. 
To model the background, we used the CALDB blank sky event files for \textit{Chandra}.
The spectral files are generated by using \texttt{specextract}.
The spectra from the observations were fitted using the XSPEC software package, version 12.9.0. \citep{Arnaud96}. 
Temperature and abundance were allowed to be free parameters. 
We adopt the solar abundance table from \citet{andgrev}. 
Varying the galactic absorption column density did not significantly influence the primary parameters or improve the spectral fits. 
Therefore in all cases, the value is fixed at A3563 position on the sky from the Leiden/Argentine/Bonn (LAB) Survey 
\citep{Kalberla} of $N_H=4.14\times10^{20}$ $cm^{-2}$.

\begin{table}
\begin{center}
\caption{\label{spect}The best-fit parameters of spectral modeling for A3653 subclusters with  $^1$\textit{XMM-Newton}  and $^2$\textit{Chandra} data.}
\begin{tabular}{@{}l c c r@{}} 
\hline
\hline
Region			& 	$kT$					& Abundance  				& $\chi^2/dof$ 			\\	
				& 	(keV) 				& ($Z_{\odot}$) 			& 	 		 		\\
\hline
A3653E$^{1}$	 	&	4.67$\pm{0.20}$		&	0.15$\pm{0.05}$		& 	2162/1760 = 1.23	\\
A3653E$^{2}$ 		& 	4.67$\pm{0.46}$ 		&  	0.24$^{+0.14}_{-0.13}$	&	289/251 = 1.15		\\
A3653W$^{1}$ 		& 	3.66$^{+0.29}_{-0.28}$  	&	0.27$^{+0.09}_{-0.08}$	&	1241/1223 = 1.01	\\
\hline
\end{tabular}
\label{table}
\end{center}
\end{table}

\begin{figure*}
 \begin{center}
 \includegraphics[width=17cm]{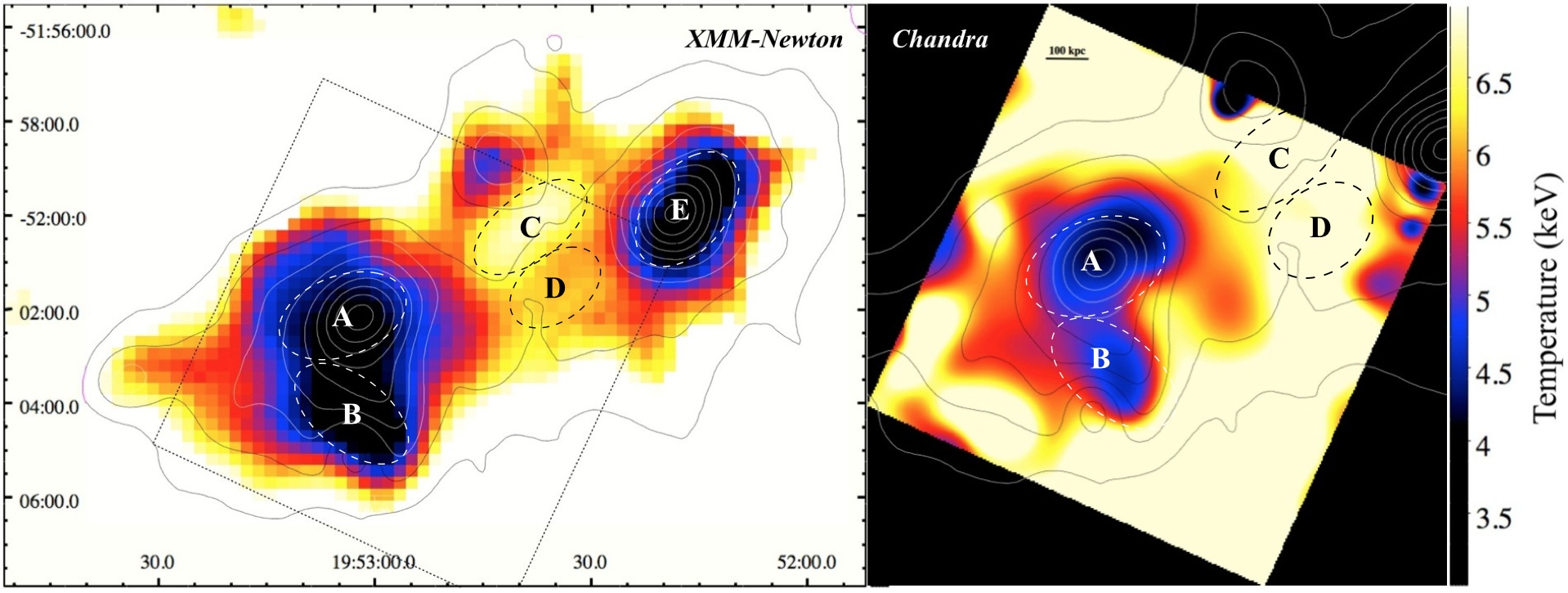}
 \caption{\label{temperature} Temperature map obtained by hardness ratio approximation method 
 for \textit{XMM-Newton} (left) and \textit{Chandra} (right) (see $\S$-\ref{temp} for detail).
 The color-coding ranges are from black ($\sim$4 keV) to yellow ($>$6 keV).
 X-ray contours (as in Fig.\ref{optic}) are superimposed for visual aid.
 The tilted-square shows \textit{Chandra} FOV. 
 The regions used for the spectroscopic analysis are displayed in ellipses.}
\end{center} \end{figure*}

For \textit{XMM-Newton} data, the energies outside the range $0.3-10$ keV were ignored.
The EPIC-MOS and PN spectra were fitted simultaneously, after checking the consistency.
The bright instrumental lines due to quiescent particle background (QDP); 
Al-K$\alpha$ (1.49 $keV$), Si-K$\alpha$ (1.74 $keV$) for MOS and Al-K$\alpha$, Cu-K$\alpha$ (8.05, 8.91 $keV$) for PN are carefully ignored from spectral data. 
The \textit{Chandra} fitting was performed in the $0.5-7$ keV energy range. 
The blank sky event files normalised at $10-12$ keV band to the count rate of the observation to account for the QDP. 
We used an absorbed single temperature collisional equilibrium plasma (APEC) model. 
For A3563E subcluster, the model gives a temperature of $kT=4.67\pm{0.20}$ keV and 
an abundance of $Z=0.14\pm{0.05}$ $Z_{\odot}$ for \textit{XMM-Newton}.
For the same region, the best-fit parameters of \textit{Chandra} data are highly consistent within the error range;
the temperature is kT$=$4.67$\pm{0.46}$ keV, and the abundance is Z=0.20$^{+0.14}_{-0.13}$ $Z_{\odot}$.
Since we lacked \textit{Chandra} observation, the spectral modeling is only performed for \textit{XMM-Newton} data for western subcluster A3563W. 
The best fit parameters for the spectral modelling within 2.5 arcmin radius circle 
are 3.66$^{+0.29}_{-0.28}$ keV of temperature and 0.27$^{+0.09}_{-0.08}$ $Z_{\odot}$ of abundance. 
The results from the spectral fitting are summarised in Table \ref{spect}. 


\begin{table}
\begin{center}
\caption{\label{templist}Comparison of the temperature map ($T_{map}$) values with spectroscopically derived temperatures ($T_{spec}$).}
\begin{tabular}{@{}cccc@{}} 
\hline
\hline
Region	& 	$T_{map}$ 		&	$T_{spec}$ 	&	$\chi^2$/dof		\\
		& 	 (keV)				&		 (keV)		&					\\
\hline
A	 	& 4.21$\pm{0.49}$		& 4.30$^{+0.34}_{-0.28}$ 	& 	1166/1122 = 1.03	\\
B	 	& 3.66$\pm{0.63}$		& 3.78$^{+0.36}_{-0.35}$ 	&	717/693 = 1.03		\\
C	 	& 6.08$\pm{0.99}$		& 6.16$^{+1.40}_{-1.20}$ 	&	373/364 = 1.02		\\
D		& 5.98$\pm{0.97}$ 		& 6.06$^{+1.20}_{-1.10}$ 	&	575/485 = 1.18		\\
E		& 3.65$\pm{0.57}$		& 3.38$^{+0.49}_{-0.53}$ 	&	240/241 = 1.00		\\
\hline
\end{tabular}
\label{table}
\end{center}
\end{table}

\subsection{\label{temp} Temperature Map}

Temperature maps are very powerful tools and provide useful information about temperature discontinuities due to the ongoing merger. 
To search for evidence, we generated a temperature map for A3653 with hardness ratio approximation. 
The images have been extracted in soft and hard energy bands carefully by avoiding QDP instrumental lines, 
galactic absorption at soft bands ($<0.7$ keV) and point source emissions at hard bands ($>8$ keV).  
Point sources excluded event files used for the analysis,
but the source excluded holes were filled using the surface brightness of the surrounding pixels and the CIAO tool \texttt{dmfilt}.
The soft and hard energy bands were selected to have approximately equal photon counts, which is also statistically favourable. 
The hardness ratio maps are produced by the images at energy bands $0.8-1.6$ keV and $1.8-8$ keV from \textit{XMM-Newton} MOSs, 
$0.7-1.6$ keV and $1.6-7$ keV for \textit{Chandra} ACIS-S3 counts, respectively. 
The hard images are divided by the corresponding soft images to obtain a hardness ratio map.  
The pixel ratio values of the map are converted to temperature values by multiplying the theoretical conversion factors. 
The factors are determined using an absorbed single thermal collisional plasma model (wabs + APEC), 
with a column density fixed to the Galactic value of $N_H=4.14\times10^{20}$ $cm^{-2}$, and the redshift at $z=0.1089$.
A similar technique for producing temperature map from hardness ratio is outlined in \citet{Ferrari06}.

Figure \ref{temperature} shows the temperature map obtained through the hardness ratio technique for \textit{XMM-Newton} (left) and \textit{Chandra} (right). 
The color$-$coding is arranged to display the similar temperature values for both, which range from black ($\sim$4 keV) to yellow ($\sim$6 keV).
The \textit{Chandra} and \textit{XMM-Newton} temperature structure and the variation display similar properties within the common frame, 
which is shown with a square in the left panel. 
Based on the temperature map, A3653 is characterized by two cold ($\sim$3.5 to 4.5 keV) regions associated with A3653E \& A3653W, 
and a hot ($\sim$6 keV) region in between. 

\begin{figure}
 \begin{center}
 \includegraphics[width=8cm]{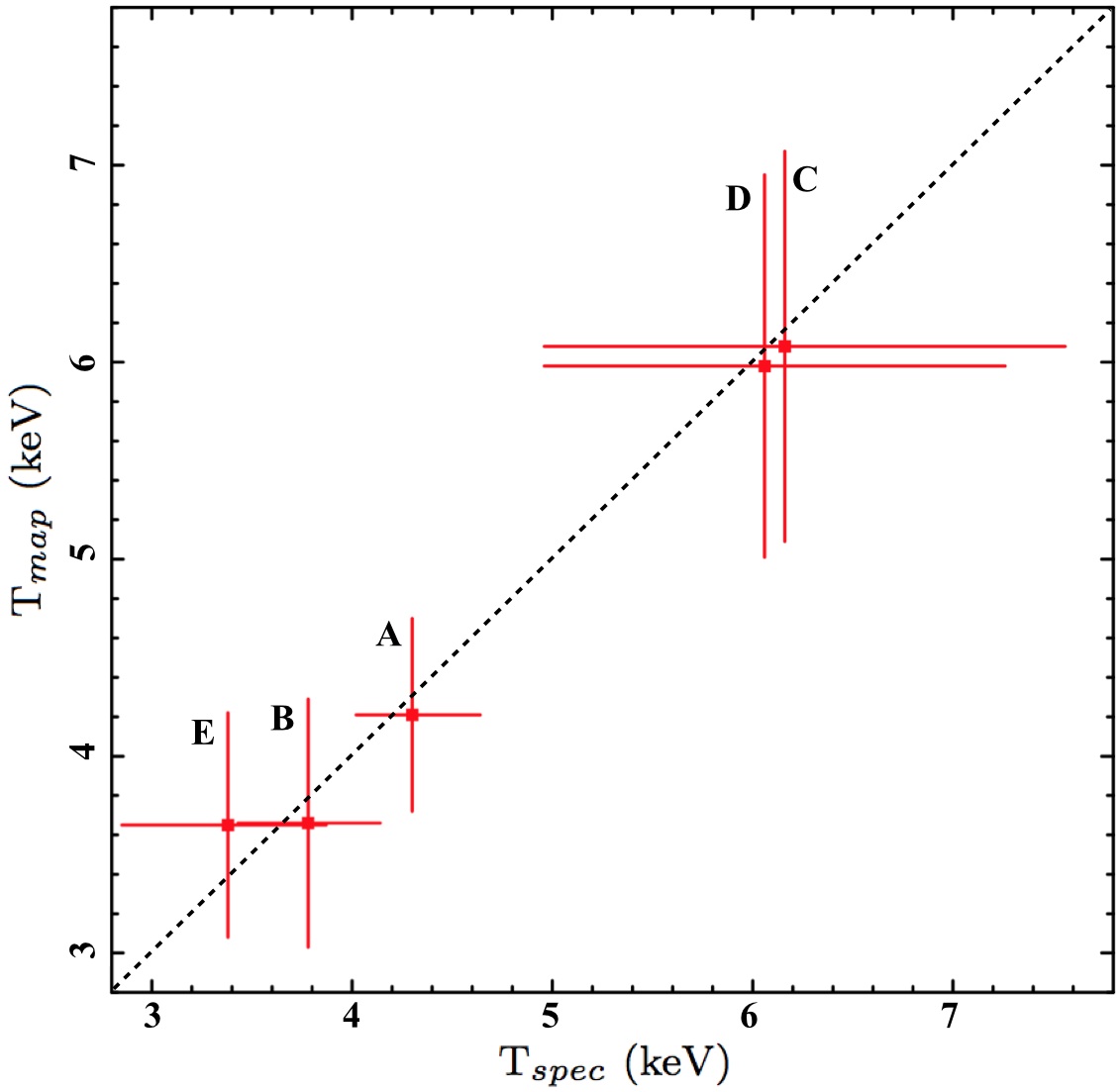}
 \caption{\label{mapspec} The plot of the $T_{spec}$ vs. $T_{map}$ for 5 peculiar regions. 
 The ratios vary around 1 (inclined line), which confirms consistency and the agreement of the values.}
\end{center} \end{figure}

The confidence of the produced temperature map is spectroscopically confirmed 
with 5 control regions (labelled A, B, C, D and E) as shown in Figure \ref{temperature}.  
The spectroscopic temperature values ($T_{spec}$) are estimated from \textit{XMM-Newton} MOSs/PN simultaneous fit 
with $N_H$ fixed to the galactic value, and provide a reduced $\chi^2\simeq1$.
Table \ref{templist} compares these best-fit temperature values ($T_{spec}$) with the values obtained directly from temperature maps ($T_{map}$). 
The values are in good agreement within the error range.
Figure \ref{mapspec} shows the plot of $T_{spec}$ vs. $T_{map}$ values and visually confirms the consistency.
The ratio values neatly vary around 1 (inclined-line), where $T_{spec}$ and $T_{map}$ are equal. 
 
Temperature map study and related spectral analysis results verify that A3653E (4.67 KeV) and A3653W (3.66 keV) has relatively cool cores.
There is a significantly hot (6.16 keV) distinct region between the subclusters. 
If we assume that the subclusters are gravitationally bound and moving towards each other, the hot region seems like shock heated by a possible merger event.

\section{Mass Analysis}
The mass of isothermal clusters is estimated to scale with the X-ray emission-weighted temperature by $M \propto T_{x}^{3/2}$ as predicted by simulations \cite[e.g.,][]{eke1996} and observations \cite[e.g.,][]{Markevitch1998}. 
A similar assumption gives us a mass ratio of 1.4:1.

The masses were also quantitatively estimated using the $M-T$ relation derived without groups by \citet[][]{Finog2001}:
\begin{equation}
    M_{500}=(3.57^{+0.41}_{-0.35})10^{13}\times kT_{ew}^{1.58^{+0.06}_{-0.07}}
\end{equation}
A3653E has a best-fit global temperature of $4.67\pm0.20$ keV, and an abundance of $0.15\pm0.05$ Z$_{\odot}$ value. 
The scaling relation estimates a total mass of $4.06\times 10^{14}$ M$_{\odot}$ for A3653E subcluster. 
The best fit temperature of A3653W is $3.66^{+0.29}_{-0.28}$ keV, with an abundance of $0.27^{+0.09}_{-0.08}$ Z$_{\odot}$.  
The scaling relation predicts a total mass of $2.77\times 10^{14}$ M$_{\odot}$ for A3653W. 
These numeric mass estimations also give a mass fraction of $\sim$1.4, consistently.
Table \ref{properties} lists redshifts, the relative velocity of BCGs and total masses of A3653 subclusters. 

\begin{table}
\begin{center}
\caption{\label{properties} Properties of A3653 system.}
\begin{tabular}{@{}lll@{}} 
\hline
\hline
Parameter					& 	A3653E				&	A3653W			\\
\hline
$z$	 					& 	0.1099$\pm$0.0002$^a$		&	0.1075$\pm$0.0002$^b$ 	\\
$V_r$ (km s$^{-1}$)			& 	32947$\pm$66$^a$			& 	32221$\pm$45$^b$		\\
$M (M_\odot$)				&	4.06$\times10^{14}$		& 2.77$\times10^{14}$	\\
\\
\hline
\end{tabular}
\label{table}
\end{center}
\footnotesize{$^a$\citet{Pimbbett}, $^b$\citet{jones2009}} 
\end{table}


\section{A Dynamical Model for A3653E \& A3653W}
We apply the Newtonian gravitational binding criterion that two-body system is bound if the potential energy of the system is equal or greater than the kinetic energy.
The two-body dynamical model was described in detail by \citet{Beers} and \citet{Greg84}.
The model was also successfully applied to several bimodal systems in the literature:
e.g. A1367 \citep{cortese}, A168 \citep{hwang}, A2319 \citep{peng}, A3716 \citep{andreda}, A3407 \& A3408 \citep{nash}
 and A1750 \citep{hwang, Bulbul}.
This model allows us to evaluate the dynamical state of A3653E and A3653W and
to estimate the probability that (i) the system infalling, (ii) the system is gravitationally bound but still expanding, 
or (iii) the subclusters are unbound but are projectionally close together in the sky by chance.
The limits of the bound solutions for a system can be determined by using the Newtonian criterion for gravitational binding \citep{Beers}:
\begin{equation}
    V_r^2 \ R_p  \leq \ 2GM \ sin^2 \alpha \ cos\alpha    
\end{equation}
The radial velocity difference, $V_r$, and the projected separation, $R_p$, are related to the real velocity and separation of the system parameters by
\begin{equation}
    V_r= V\ sin\alpha, \ R_p={R} \ cos\alpha.
\end{equation}
The angle $\alpha$ is the projection angle between the plane of the sky and the line connecting the subsystems 
(i.e., $\alpha$=0 if the subclusters are at the same distance). 
$V$ and $R$ are true (3D) velocity and positional separation between the two subclusters. 
For the gravitationally bound systems the parametric solutions to the equation of motion \citep{Beers} :
\begin{equation}
    t=\bigg(\frac{R_m^3}{8GM}\bigg)^{1/2}   (\chi-sin \ \chi),
\end{equation}
\begin{equation}
    R=\frac{R_m}{2}(1-cos \ \chi),
\end{equation}
\begin{equation}
    V=\bigg(\frac{2GM}{R_m}\bigg)^{1/2}    \frac{sin \ \chi}{(1-cos \ \chi)},
\end{equation}
where $R$ is the separation at time $t$, $R_m$ is the separation at the maximum expansion, 
$M$ is the combined mass of the system, and $\chi$ is the developmental angle. 
Similarly, the parametric solutions for the unbound case is also described by \citet{Beers}.
For gravitationally unbound systems, the parametric equations are
\begin{equation}
    t= \frac{GM}{V_\infty^3}   (sinh \ \chi - \chi),
\end{equation}
\begin{equation}
    R=\frac{GM}{V_\infty^2}(cosh \ \chi -1),
\end{equation}
\begin{equation}
    V=  V_\infty   \frac{sinh \ \chi}{(cosh \ \chi -1)},
\end{equation} 
where $V_\infty$ is the asymptotic expansion velocity. 
Using the parameters previously found for A3653 subclusters: 
radial velocity difference of $V_r$ = 726 $\pm$ 80 km s$^{-1}$, 
the projected distance ($R_p$) on the plane of the sky between the X-ray centers of the subclusters as 0.89 Mpc, 
total mass of $M=6.83\times10^{14}$ (see Table \ref{properties}), and 
setting $t=12.1$ Gyr (3.8$\times10^{17} s$) the age of the universe at A3653 redshift, 
we can constrain the equations. 
Subsequently, the radial velocity difference ($V_r$)  as a function of the projection angle ($\alpha$)
between the subclusters are solved by equation (6) of \citet{Greg84} for bound and unbound states are
\begin{equation}
  tan \ \alpha=\frac{t V_r}{R_p}   \frac {(cos \ \chi -1)^2}{sin \ \chi (\chi-sin \ \chi) },
\end{equation}
\begin{equation}
  tan \ \alpha=\frac{t V_r}{R_p}   \frac {(cosh \ \chi -1)^2}{sinh \ \chi (sinh \ \chi - \chi) },
\end{equation}
respectively.

\begin{figure}
 \begin{center}
 \includegraphics[width=8cm]{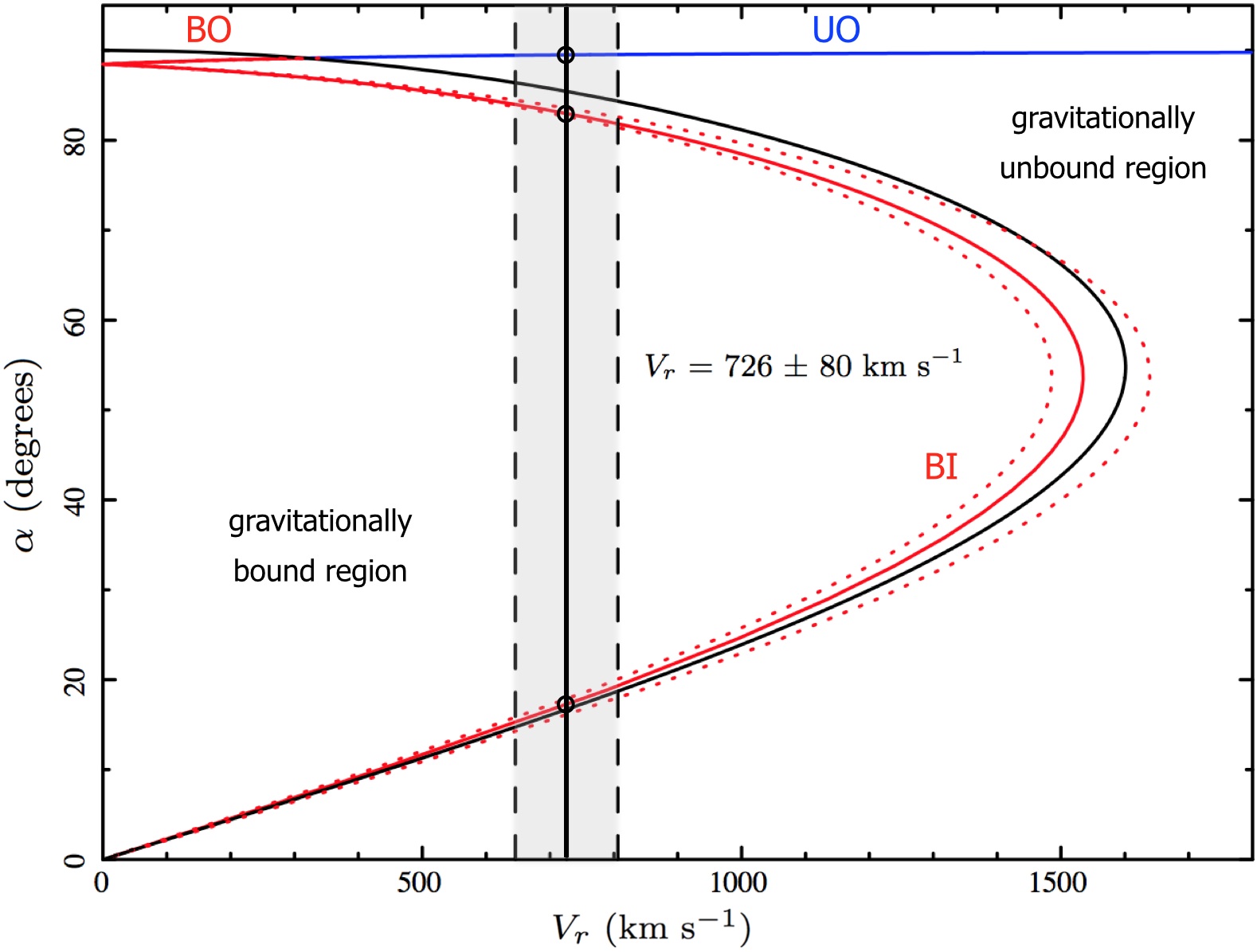}
 \caption{\label{dynamic} Projection angle ($\alpha$) 
as a function of the relative radial velocity difference ($V_r$) of the subclusters A3653E and A3653W.
BI, BO and UO stand for Bound-Incoming, Bound-Outgoing, and Unbound-Outgoing solutions. 
Red and blue lines correspond to bound and unbound solutions, respectively. 
The black curve separates the bound and unbound regions according to the Newtonian criterion. 
The vertical lines represent the relative radial velocity, $V_r$ = 726 $\pm$ 80 km s$^{-1}$, between the two clusters in the rest frame of A3653.
The black circles are the three acceptable solutions.}
\end{center} \end{figure}

The solution for A3653 system is shown in Figure \ref{dynamic}.
The projected angle ($\alpha$) as a function of the radial velocity difference ($V_r$) is plotted for bound (red line) and unbound (blue line) solutions. 
The black curve separates the bound and unbound regions according to the Newtonian criterion. 
Since the relative radial velocity of our system is observed to be $V_r$ = 726 $\pm$ 80 km s$^{-1}$, 
we obtain two bound solutions and one unbound solution.
All three solutions (black circles in Fig-\ref{dynamic}) are defined by the intersections of curved lines with 
the solid vertical line corresponding the relative velocity difference ($V_r$ = 726 $\pm$ 80 km s$^{-1}$) between the two clusters in the rest frame of A3653.
The uncertainties in the measured velocity lead to a range of solutions from $\alpha_{inf}$ to $\alpha_{sup}$ for the projection angles.
We compute relative probabilities for the solutions by the formula \citep{Girardi}:
\begin{equation}
    p_i = \int_{\alpha_{inf,i}}^{\alpha_{sup,i}}  \ cos\alpha \ d\alpha    
\end{equation}
where each solution is represented by index $i$. 
The probabilities are normalised by $P_i=p_i/(\sum_i p_i)$.

Solving the parametric equations we get two gravitationally bound incoming solutions and one unbound outgoing solution.
For the bound solutions, the subclusters are either approaching to each other at 732 km s$^{-1}$ (6.5$\%$ probability) or at 2400 km s$^{-1}$ (93.5$\%$ probability).
The first solution suggest that subcluster cores will cross each other after a long time (9.7 Gyr) given a separation of 7.24 Mpc with a mean colliding velocity of 732 km s$^{-1}$.
This bound scenario apparently does not predict a strong interaction between A3653E and A3653W, thus the solution has a low probability (6.5\%).
The latter more likely scenario corresponds to a collision in about 380 Myr, considering their separation of 0.93 Mpc with a supersonic colliding velocity of 2400 km s$^{-1}$.
Temperature map (Fig-\ref{mapspec}) also confirms a hot region and spectral fit gives kT= 6.16$^{+1.40}_{-1.20}$ keV, providing evidence of shock heated gas between the subclusters.
The large angle ($\alpha$=89$^\circ$.49) unbound solution (0.05\% probability) corresponds to a separation of 71.3 Mpc.
The parameters of these solutions are presented in Tables \ref{solution} and \ref{unbsol}. 
Given its negligibly low probability, the unbound solution can be disregarded, 
while the bound solution close to the plane of sky ($\alpha$=17$^\circ$.61) is highly favoured (93.5\% probability).

The dynamical model analysis result for A3653 is fairly conclusive that  the subclusters are very likely gravitationally bound, 
and the cores will cross each other in 380 Myr. 
Normally, approaching systems produce strong ram pressure and displace X-ray plasma \citep{vr2013}.
In our case, X-ray peaks are found to be in the opposite direction. 
This suggests that it is not the first encounter of the subclusters; 
therefore, there was a previous core passage.
The probabilities for the bound solutions should be treated cautiously, 
since the dynamical two-body model assumes a clear radial infall (e.g. head-on merger) and 
does not include the angular momentum information of the subclusters, which is very unlikely to be zero.
If the merging between the subclusters occurs with angular momentum as in off-axis mergers, consequences will be more complicated as shown in the simulations \citep{taki, vr2013}.


\begin{table}
\begin{center}
\caption{\label{solution} Best-Fit Parameters for the Bound Incoming Solutions of the Dynamical Model.
The columns list best-fit values for $\chi$ and $\alpha$ for the bound solutions, 
and the corresponding values for $R$, $R_m$, $V$ and the related probabilities $P$.}
\begin{tabular}{@{}cccccc@{}} 
\hline
\hline
 $\chi$	& 	$\alpha$	&	$R$ 		&	$R_m$	& 	$V$			&	$P$	\\	
 (rad)	&	(degrees)	&	(Mpc)	&	(Mpc)	&	(km s$^{-1}$)	&	(\%) 	\\
\hline
5.05 		&	82.94	& 	7.24		&	21.50	&	731.5 		&	6.5 \\
5.86		&	17.61	& 	0.93		&	20.88	&	2399.7		&	93.5 \\
\hline
\hline
\end{tabular}
\label{table}
\end{center}
\end{table}

\begin{table}
\begin{center}
\caption{\label{unbsol} Best-Fit Parameters for the Unbound Outgoing Solutions of the Dynamical Model.
The columns list best-fit values for $\chi$ and $\alpha$ for the bound solutions, 
and the corresponding values for $R$, $V$, $V_\infty$ and the related probabilities $P$.}
\begin{tabular}{@{}cccccc@{}} 
\hline
\hline
 $\chi$	& 	$\alpha$	&	$R$ 		& 	$V$			&	$V_\infty$		&	$P$	\\	
 (rad)	&	(degrees)	&	(Mpc)	&	(km s$^{-1}$)	&	(km s$^{-1}$)	&	(\%) 	\\
\hline
3.16 		&	89.49	& 	71.30	&	726.03 		&	666.88		&	0.02 \\
\hline
\hline
\end{tabular}
\label{table}
\end{center}
\end{table}

\section{Summary}

We present the analysis results of a merging binary cluster A3653 using \textit{AAT} Optical Galaxy Survey Data, 
\textit{XMM-Newton} and \textit{Chandra} X-ray observations. 
Spectroscopic redshift analysis, X-ray brightness \& temperature maps and two-body dynamical model analysis
clearly indicate that A3653 is not a single structure but composed of two subclusters. 
A possible interpretation of our findings is illustrated in Figure \ref{sum}.
Our main results are:

\begin{figure}
 \begin{center}
 \includegraphics[width=8cm]{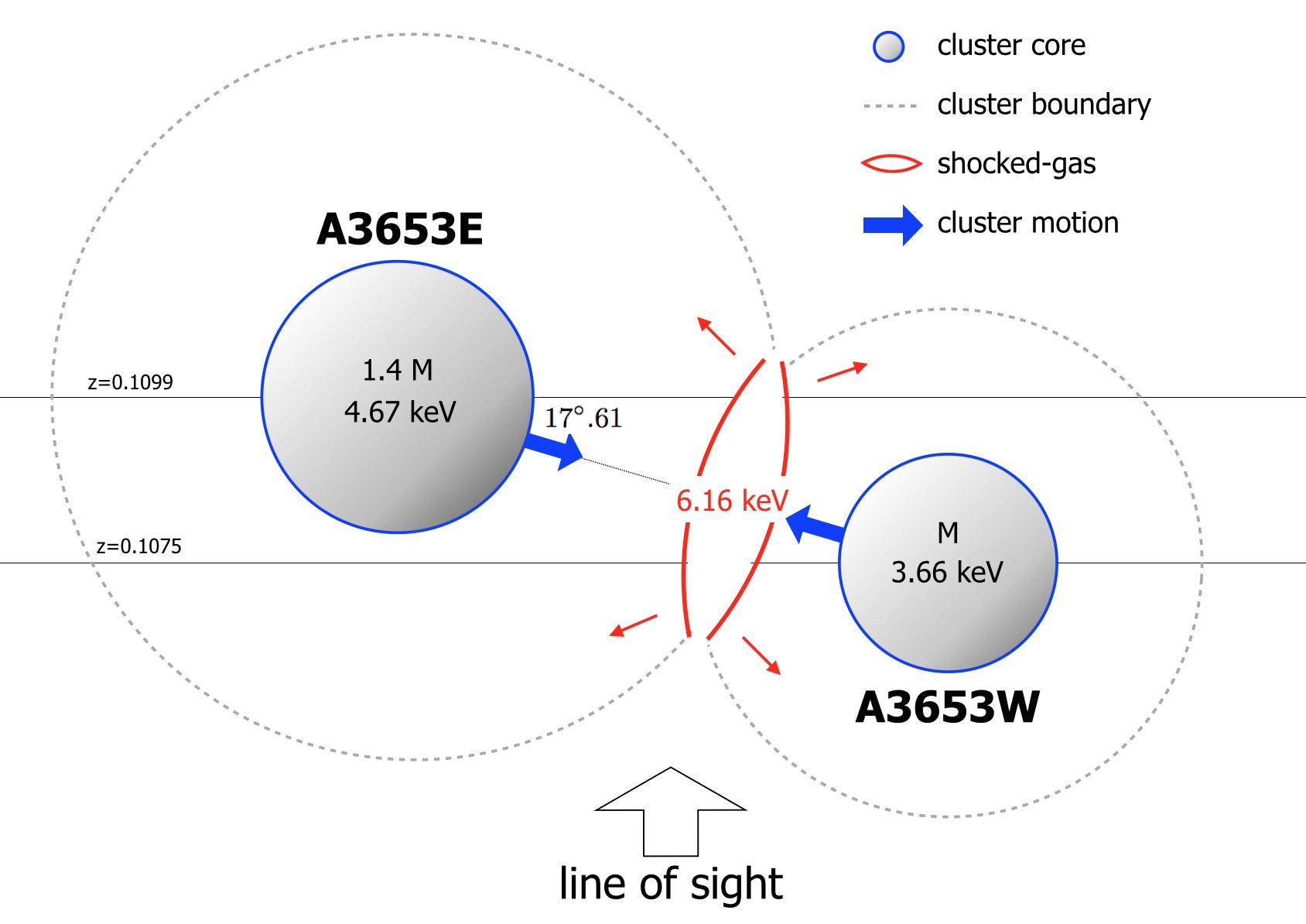}
 \caption{\label{sum}Tentative illustration of A3653E \& A3653W subclusters gravitationally falling into each other (blue-arrows).
The merger is happening very close to the plane of sky (horizontal-lines). 
The cluster boundaries (dotted-circles) start to collide and develop a shocked region (in red).
The line of sight is depicted with big arrow at the bottom.}
\end{center} \end{figure}

\begin{itemize}
\item On the basis of the optical spectroscopic redshifts of 87 member galaxies with AAT Galaxy Survey data, A3563 has a mean cluster redshift value of $z=0.1089$.
We detect two dominant grouping of member galaxies; BCG1 at z=0.1099, can be associated with the main cluster and a little foreground western subgroup concentrated at z=0.1075, 
BCG2 (see Fig-\ref{opt_density}).

\item X-ray brightness map also clearly indicates the binary structure of A3653 with two distinct clumps (see Fig-\ref{raw_image}).
The X-ray peaks match with BCGs by a small disturbance ($\sim$35 kpc), 
which is probably caused by ram-pressure of an ongoing dynamical activity (see Fig-\ref{optic}). 
The displacement of the peaks aligns with the elongation of the subcluster locations, 
which suggest that the peak shifts are associated with the merger activity. 
However, the position of the peaks imply that there was a previous core passage between the subclusters.

\item Average temperature values of A3563E and A3563W are 4.67 keV and 3.66 keV, respectively. 
Based on the temperature map, we detect a possible shock heated gas with a significantly elevated temperature of $kT=$ 6.16 keV between the two subclusters (see Fig-\ref{temperature}).  
The hot gas results from shock heating by merging subclusters.

\item The mass calculations for isothermal clusters result in $M_{A3653E}=4.06\times 10^{14}$ M$_{\odot}$ and $M_{A3653W}=2.77\times 10^{14}$ M$_{\odot}$.
A3653 is composed of two merging subclusters with $\sim$1.4 mass fraction. 

\item The two-body dynamical model suggests three solutions: two collapsing and one unbound expanding (see Fig-\ref{dynamic}). 
Based on the analysis, the unbound solution has negligibly low (0.02\%) probability. 
The subclusters are gravitationally bound and infalling. 
The merging is happening close to the plane of sky ($\alpha$=17$^\circ$.61), and the cores will cross each other in 380 Myr.

\end{itemize}

\section*{Acknowledgement}

We would like to acknowledge financial support from the Scientific and Technological Research Council of Turkey (T\"{U}B\.{I}TAK) project number 113F117. 
The work is also supported by YTU Scientific Research \& Project Office (BAP) funding with contract numbers 2013-01-01-YL01 and 2013-01-01-KAP04. 
The authors would like to thank S. Alis for the useful discussion and suggestions.

\begin{table*}
\footnotesize
\begin{center}
\small
\caption{List of member galaxies of Abell 3653. The galaxies numbered 19, 22, 28 and 71 are from 6dF survey.}
\begin{tabular}{r l c ccc c}
\hline
\#  	& Object Name					&	RA     	& Dec     		& Velocity		&	Redshift  	&  Magnitude  \\
 	& 							&	(J2000)    	& (J2000)    	& km/s  		&   			& Filter \\
\hline
     1	& GALEXASC J195249.91-520139.6	& 298.20825	& -52.02811	&  31928	&	0.1065  & 	17.5R \\
     2	& ABELL 3653:PRD 351			& 298.25292	& -52.02064	&  32587	&	0.1087  & 	17.7R \\
     3	&2MASX J19525707-5202462		& 298.23787	& -52.04611	&  32887	&	0.1097  & 	16.2R \\
     4	&	2MASX J19530336-5202132	& 298.26404	& -52.03700	&  32947	&	0.1099  & 	11.2R \\
     5	&	ABELL 3653:PRD 133		& 298.27417	& -52.02500	&  32198	&	0.1074  & 	18.4R \\
     6	&	ABELL 3653:PRD 347		& 298.22688	& -52.05733	&  32737	&	0.1092  & 	17.4R \\
     7	&	ABELL 3653:PRD 285		& 298.27729	& -52.03906	&  32947	&	0.1099  & 	18.1R \\
     8	&	ABELL 3653:PRD 010		& 298.18763	& -51.99322	&  31298	&	0.1044  & 	17.2R \\
     9	&	ABELL 3653:PRD 349		& 298.23954	& -51.97142	&  31628	&	0.1055  & 	17.9R \\
    10	&	ABELL 3653:PRD 018		& 298.13488	& -52.00225	&  32108	&	0.1071  & 	18.1R \\
    11	&GALEXASC J195313.18-520313.3	& 298.30600	& -52.05344	&  32617	&	0.1088  & 	18.2R \\
    12	&	ABELL 3653:PRD 050		& 298.16692	& -52.07353	&  32977	&	0.1100  & 	18.2R \\
    13	&	ABELL 3653:PRD 101		& 298.12083	& -52.02139	&  32258	&	0.1076  & 	 17.7R \\
    14	&	ABELL 3653:PRD 035		& 298.10217	& -52.04547	&  30549	&	0.1019  & 	 18.7R \\
    15	&GALEXASC J195239.36-520549.8	& 298.16367	& -52.09700	&  31868	&	0.1063  & 	 17.7R \\
    16	&GALEXASC J195306.88-515647.9	& 298.27804	& -51.94678	&  31478	&	0.1050  & 	 17.1R \\
    17	&	ABELL 3653:PRD 338		& 298.12354	& -52.08539	&  31868	&	0.1063  & 	 17.7R \\
    18	&	ABELL 3653:PRD 245		& 298.08171	& -52.02192	&  31598	&	0.1054  & 	 17.6R \\
    19	&	2MASX J19521735-5159465	& 298.07238	& -51.99622	&  32221	&	0.1075  & 	 16.1b \\
    20	&	2MASX J19531237-5206322	& 298.30163	& -52.10883	&  31178	&	0.1040  & 	16.9R \\
    21	&	ABELL 3653:PRD 230		& 298.06113	& -52.00808	&  32198	&	0.1074  & 	17.8R \\
    22	&	2MASX J19521995-5205095	& 298.08312	& -52.08589	&  31604	&	0.1054  & 	15.9R \\
    23	&GALEXASC J195211.95-520042.3	& 298.04983	& -52.01147	&  29769	&	0.0993  & 	17.6R \\
    24	&	ABELL 3653:PRD 029		& 298.04363	& -52.00411	&  32348	&	0.1079  & 	17.9R \\
    25	&	ABELL 3653:PRD 006		& 298.31296	& -51.92744	&  31298	&	0.1044  & 	18.7R \\
    26	&	ABELL 3653:PRD 358		& 298.38788	& -51.97586	&  32468	&	0.1083  & 	17.8R \\
    27	&GALEXASC J195308.09-515501.9	& 298.28367	& -51.91731	&  32827	&	0.1095  & 	18.0R \\
    28	&	2MASX J19531839-5155491	& 298.32646	& -51.93039	&  30836	&	0.1029  & 	16.1R \\
    29	&GALEXASC J195209.58-520339.2	& 298.03967	& -52.06108	&  31658	&	0.1056  & 	17.3R \\
    30	&	ABELL 3653:PRD 340		& 298.14100	& -51.91719	&  32468	&	0.1083  & 	18.1R \\
    31	&	ABELL 3653:PRD 057		& 298.35767	& -52.11078	&  30999	&	0.1034  & 	17.9R \\
    32	&	2MASX J19531914-5155201	& 298.3300	& -51.92233	&  31298	&	0.1044  & 	16.4R \\
    33	&	ABELL 3653:PRD 334		& 298.04671	& -51.96708	&  31178	&	0.1040  & 	17.9R \\
    34	&GALEXASC J195336.19-515744.5	& 298.39996	& -51.96253	&  30279	&	0.1010  & 	17.9R \\
    35	&	ABELL 3653:PRD 198		& 298.25192	& -52.15142	&  31568	&	0.1053  & 	18.8R \\
    36	&	ABELL 3653:PRD 362		& 298.41796	& -52.06881	&  33727	&	0.1125  & 	17.8R \\
    37	&	ABELL 3653:PRD 172		& 298.01963	& -51.98944	&  31658	&	0.1056  & 	18.0R \\
    38	&	ABELL 3653:PRD 049		& 298.24442	& -52.15547	&  31868	&	0.1063  & 	18.3R \\
    39	&GALEXASC J195326.15-520801.2	& 298.35921	& -52.13342	&  34386	&	0.1147  & 	19.1R \\
    40	&	2MASX J19531989-5154211	& 298.33308	& -51.90594	&  32917	&	0.1098  & 	17.0R \\
    41	&	2MASX J19515849-5203084	& 297.99375	& -52.05239	&  32767	&	0.1093  & 	16.2R \\
    42	&	ABELL 3653:PRD 332		& 297.99037	& -52.00719	&  31418	&	0.1048  & 	18.3R \\
    43	&	ABELL 3653:PRD 007		& 298.37521	& -51.91339	&  32917	&	0.1098  & 	18.3R \\
    44	&	ABELL 3653:PRD 282		& 298.44896	& -52.07464	&  32737	&	0.1092  & 	18.0R \\
    45	&	ABELL 3653:PRD 130		& 298.39583	& -52.13614	&  31718	&	0.1058  & 	17.1R \\
    46	&	ABELL 3653:PRD 364		& 298.42821	& -52.11667	&  32138	&	0.1072  & 	17.2R \\
    47	&GALEXASC J195304.51-515211.0	& 298.26929	& -51.86925	&  31928	&	0.1065  & 	17.4R \\
    48	&	ABELL 3653:PRD 331		& 297.97608	& -51.96444	&  32498	&	0.1084  & 	17.7R \\
    49	&	2MASX J19530945-5211122	& 298.28933	& -52.18678	&  33187	&	0.1107  & 	16.7R \\
    50	&	ABELL 3653:PRD 090		& 298.01667	& -51.91583	&  32587	&	0.1087  & 	17.4R \\
    51	&GALEXASC J195254.41-515127.2	& 298.22683	& -51.85711	&  30699	&	0.1024  & 	18.5R \\
    52	&GALEXASC J195155.56-520639.7	& 297.98146	& -52.11133	&  32647	&	0.1089  & 	18.0R \\
    53	&	ABELL 3653:PRD 279		& 298.49304	& -52.06967	&  32677	&	0.1090  & 	18.2R \\
    54	&	2MASX J19535592-5205196	& 298.48292	& -52.08881	&  31328	&	0.1045  & 	16.4R \\
    55	&	ABELL 3653:PRD 328		& 297.94571	& -51.97167	&  32348	&	0.1079  & 	18.4R \\
    56	&GALEXASC J195226.88-515104.1	& 298.11175	& -51.85139	&  30729	&	0.1025  & 	17.3R \\
    57	&	ABELL 3653:PRD 093		& 297.96383	& -51.92294	&  33397	&	0.1114  & 	18.1R \\
    58	&GALEXASC J195139.34-520231.9	& 297.91392	& -52.04194	&  32887	&	0.1097  & 	17.2R \\
    59	&	ABELL 3653:PRD 066		& 298.52496	& -52.07533	&  33067	&	0.1103  & 	17.7R \\
    60	&	ABELL 3653:PRD 026		& 297.93000	& -51.94375	&  32917	&	0.1098  & 	17.6R \\
    \hline
\end{tabular}
\end{center}
\end{table*}

\begin{table*}
\setcounter{table}{6}
\begin{center}
\caption{continued.}
\begin{tabular}{r l c ccc c}
\hline
\#  	& Object Name					&	RA     	& Dec     		& Velocity		&	Redshift  	&  Magnitude  \\
 	& 							&	(J2000)    	& (J2000)    	& km/s  		&   			& Filter \\
\hline
    61	&GALEXASC J195341.68-515207.5	& 298.42367	& -51.86858	&	33157	&	0.1106  & 	18.2R \\
    62	&	ABELL 3653:PRD 121		& 298.17492	& -52.22450	&	32468	&	0.1083  & 	18.8R \\
    63	&GALEXASC J195412.11-520133.6	& 298.55071	& -52.02583	&	33157	&	0.1106  & 	17.6R \\
    64	&	2MASX J19513315-5202553	& 297.88796	& -52.04878	&	32198	&	0.1074  & 	16.9R \\
    65	&	ABELL 3653:PRD 003		& 298.53050	& -51.93383	&	30999	&	0.1034  & 	18.2R \\
    66	&	2MASX J19532263-5213301	& 298.34417	& -52.22506	&	32587	&	0.1087  & 	16.5R \\
    67	&GALEXASC J195301.00-514835.7	& 298.25417	& -51.80981	&	33697	&	0.1124  & 	17.6R \\
    68	&	ABELL 3653:PRD 124		& 298.38650	& -52.21856	&	32318	&	0.1078  & 	18.5R \\
    69	&	ABELL 3653:PRD 186		& 298.36450	& -52.22814	&	32348	&	0.1079  & 	18.2R \\
    70	&	ABELL 3653:PRD 325		& 297.91596	& -52.14275	&	32018	&	0.1068  & 	18.4R \\
    71	&	2MASX J19541986-5158230	& 298.58279	& -51.97308	&	33162	&	0.1106  & 	15.9R \\
    72	&	2MASX J19542245-5200250	& 298.59346	& -52.00694	&	30219	&	0.1008  & 	16.6R \\
    73	&	ABELL 3653:PRD 140		& 298.59604	& -52.03164	&	33847	&	0.1129  & 	18.1R \\
    74	&GALEXASC J195355.47-521137.6	& 298.48092	& -52.19364	&	33847	&	0.1129  & 	16.9R \\
    75	&	ABELL 3653:PRD 031		& 297.84621	& -52.04850	&	32258	&	0.1076  & 	17.9R \\
    76	&	ABELL 3653:PRD 120		& 298.11296	& -52.25203	&	32108	&	0.1071  & 	17.9R \\
    77	&GALEXASC J195253.38-514642.8	& 298.22225	& -51.77856	&	31478	&	0.1050  & 	19.4R \\
    78	&	ABELL 3653:PRD 080		& 298.41883	& -51.80536	&	32887	&	0.1097  & 	17.4R \\
    79	&GALEXASC J195117.03-520619.7	& 297.81967	& -52.10514	&	32378	&	0.1080  & 	18.7R \\
    80	&	2MASX J19532972-5216226	& 298.37396	& -52.27300	&	32767	&	0.1093  & 	17.0R \\
    81	&	ABELL 3653:PRD 141		& 298.65158	& -51.99289	&	31478	&	0.1050  & 	18.2R \\
    82	&	ABELL 3653:PRD 181		& 297.96808	& -52.24167	&	31748	&	0.1059  & 	18.6R \\
    83	&	ABELL 3653:PRD 135		& 298.66808	& -52.05789	&	32498	&	0.1084  & 	18.0R \\
    84	&	2MASX J19515019-5148042	& 297.95904	& -51.80125	&	30609	&	0.1021  & 	17.4R \\
    85	&	2MASX J19543069-5209140	& 298.62792	& -52.15383	&	31838	&	0.1062  & 	16.5R \\
    86	&	ABELL 3653:PRD 051		& 298.34125	& -52.29897	&	32647	&	0.1089  & 	17.8R \\
    87	&	ABELL 3653:PRD 187		& 298.51992	& -52.26153	&	31748	&	0.1059  & 	18.9R \\
\hline
\end{tabular}
\end{center}
\label{tab:cat}
\end{table*}

\bsp	
\label{lastpage}
\end{document}